\documentclass[showpacs,aps,twocolumn,prd,longbibliography]{revtex4-2}
\usepackage{epsfig}
\usepackage{graphicx}
\usepackage{amsmath,amssymb,amsfonts}
\usepackage{array}
\usepackage{url}
\usepackage{multirow}
\usepackage{float}
\usepackage{lineno}
\usepackage{xspace}
\usepackage{ulem}
\usepackage[usenames,dvipsnames]{color}
\definecolor{darkblue}{RGB}{0,0,196}
\usepackage[colorlinks=true,linkcolor=darkblue,citecolor=darkblue,urlcolor=darkblue]{hyperref}
\usepackage{cleveref}

\newcommand{\muB}{\mu_{\mathrm{B}}}
\newcommand{\mucapB}{\hat{\mu}_{\mathrm{B}}}
\newcommand{\mud}{m_{\mathrm{u/d}}}
\newcommand{\ms}{m_{\mathrm{s}}}
\newcommand{\mg}{m_{\mathrm{g}}}
\newcommand{\cs}{c_{\mathrm{s}}^{2}}
\newcommand{\Cv}{C_{\mathrm{V}}}
\newcommand{\TbyTc}{T/T_{\mathrm{c}}}
\begin{document}
\title{Thermodynamic and transport properties of the quark-gluon plasma \\
at finite chemical potential with a deep neural network framework}

\author{Rishabh Kumar Tiwari$^1$}
\author{Kangkan Goswami$^1$}
\author{Suraj Prasad$^{1,2}$}
\author{Captain R. Singh$^1$}
\author{Raghunath Sahoo$^1$}
\email{Corresponding Author Email: {Raghunath.Sahoo@cern.ch}}
\author{Mohammad Yousuf Jamal$^3$}
\affiliation{$^1$Department of Physics, Indian Institute of Technology Indore,
Simrol, Indore 453552, India}
\affiliation{$^2$HUN-REN Wigner Research Center for Physics, Budapest 1121, Hungary}
\affiliation{$^3$Key Laboratory of Quark and Lepton Physics (MOE) \& Institute of Particle Physics,
Central China Normal University, Wuhan 430079, China}

\begin{abstract}
The characteristics of a thermal system depend strongly on its response to thermal gradients and the underlying microscopic interactions among constituents. In the present study, we investigate the thermodynamic and transport properties of the quark-gluon plasma (QGP) at finite baryon chemical potential within a deep-learning-assisted quasi-particle model (DLQPM). The temperature ($\mathrm{T}$) and baryon chemical potential ($\muB$)-dependent thermal masses of quasi-particles are estimated using neural networks trained to reproduce lattice QCD (lQCD) results for the equation of state, obtained via a Taylor-like expansion around vanishing baryon chemical potential. The trained model acts as an effective emulator, enabling us to estimate the thermodynamic and transport properties at finite $\muB$. We compute the speed of sound, specific heat, viscosity, and conductivity of the deconfined medium. Our findings are in good agreement with available lattice calculations and other phenomenological models. The present study demonstrates that a DNN-based approach provides an efficient framework for studying the properties of the QGP at finite baryon density.
\end{abstract}

\date{\today}

\maketitle
\section{Introduction}
Understanding the properties of strongly interacting matter at finite temperature and baryon density remains one of the major challenges in quantum chromodynamics (QCD). QCD predicts that at sufficiently high temperatures and/or baryon densities, hadronic matter undergoes a transition to a deconfined phase of quarks and gluons, known as the quark–gluon plasma (QGP)~\cite{Singh:1992sp}. First-principle calculations based on lattice QCD (lQCD) provide a precise determination of the equation of state of the strongly interacting matter at vanishing baryon chemical potential~\cite{Wilson:1974sk, Creutz:1980zw, Kogut:1974ag}. However, extending these calculations to finite baryon density remains a major challenge due to the well-known fermion sign problem, which prevents direct Monte Carlo simulations in this regime~\cite{Muroya:2003qs, deForcrand:2009zkb}. This limitation poses a major obstacle for quantitative studies of QCD matter relevant to relativistic heavy-ion collisions at finite baryon density and to the physics of compact astronomical bodies like neutron stars~\cite{Glendenning:1997wn,Annala:2019puf}.\\

To overcome these limitations, several complementary approaches have been developed. Lattice calculations based on Taylor expansions around vanishing chemical potential provide controlled information at small baryon densities~\cite{Allton:2002zi, Borsanyi:2012cr}. In parallel, a variety of phenomenological models-including the Nambu–Jona-Lasinio (NJL) model~\cite{Nambu:1961tp, Nambu:1961fr, Marty:2013ita, Dwibedi:2025bdd}, the Polyakov-loop extended NJL model~\cite{Fukushima:2003fw, Ratti:2005jh, Costa:2008dp, Goswami:2023eol}, and quasi-particle models~\cite{Peshier:2000, Plumari:2011mk, Sambataro:2024mkr, Singh:2025geq} have been employed to explore the thermodynamic and transport properties of the QCD medium at finite baryon chemical potential,  $\mu_{\rm B}$. While these approaches provide valuable insights, their predictive power is often limited by model assumptions, particularly in the treatment of effective thermal masses of quarks and gluons. For example, Ref.~\cite{Srivastava:2010xa} employs an effective parametrized thermal mass for quarks and gluons, whereas the Nambu–Jona-Lasinio model~\cite{Nambu:1961tp} estimates the quark thermal mass using a gap equation motivated by an analogy with superconductivity. These assumptions lead to different treatments of thermal masses, which subsequently propagate to the final estimates of thermodynamic and transport properties.\\

Recent advances in machine learning (ML) offer new opportunities to address complex inference problems in theoretical and experimental high-energy physics, particularly in regimes where first-principles calculations are computationally prohibitive or analytically intractable. Over the past decade, ML techniques have become increasingly integrated into the analysis of collider data, event and track reconstruction, and theoretical modeling~\cite{Guest:2018yhq, Mallick:2021wop, Goswami:2024xrx, Prasad:2023zdd, Sahoo:2025vip, Larkoski:2017jix}. In particular, deep neural networks (DNNs) have demonstrated remarkable capability in learning highly non-linear and higher-dimensional mappings between physical observables and underlying model parameters~\cite{Mallick:2022alr, Mallick:2023vgi, Hornik:1989yye, LeCun:2015pmr, Baldi:2014kfa}. These features make ML especially useful in situations where traditional approaches become difficult to apply, such as in the study of strongly interacting matter. Conventional phenomenological models, including quasi-particle descriptions of the QGP, often rely on simplified parametrizations that may not fully capture the behavior observed in lQCD calculations. Motivated by this, recent studies have started to incorporate machine learning into such frameworks. For example, the deep-learning-assisted quasi-particle model, which is proposed in Refs.~\cite{Li:2022ozl, Li:2025csc, Jamal:2025gjy}, where neural networks are trained on lQCD results at vanishing baryon chemical potential to infer effective quasi-particle masses. These approaches highlight the potential of ML to provide data-driven, non-parametric representations of QCD matter; however, important challenges still remain, particularly in extending such ML frameworks to finite baryon density and ensuring consistency with underlying theoretical constraints. \\

In this work, we develop a machine-learning model that incorporates lQCD constraints at both vanishing and finite baryon chemical potential. The training dataset includes information from Taylor-expansion coefficients around $\muB = 0$~\cite{Borsanyi:2021sxv}, allowing the model to learn the dependence of thermodynamic observables on baryon density. In addition, physically motivated constraints inspired by quasi-particle models are imposed to guide the learning process. Within this framework, the neural network is trained to predict the effective quasi-particle masses with explicit temperature and baryon chemical potential dependence. These masses are subsequently used to compute thermodynamic and transport properties of the QCD medium. The predictive capability of the model is assessed through systematic comparisons with lQCD results, established phenomenological approaches, and observables inferred from Bayesian analyses of heavy-ion collision data. These results highlight the potential of machine-learning-assisted quasi-particle models as a robust and flexible framework for exploring QCD matter at finite baryon density.\\

The remainder of this paper is organized as follows. In Sec.~\ref{Formalism} we present the theoretical framework used to compute the thermodynamic and transport observables. Section~\ref{Methodology} describes the machine-learning methodology employed to determine the quasi-particle masses. The resulting predictions are presented and discussed in Sec.~\ref{Results}. Finally, Sec.~\ref{Summary} summarizes the main conclusions of this work.

\section{Formalism}
\label{Formalism}
\subsection{Quasi-particle model}
The quasi-particle model is a phenomenological framework employed to describe the behavior of QGP in the non-perturbative regime of QCD, where the relevant degrees of freedom are the thermal excitations of interacting quarks and gluons~\cite{Goloviznin:1992ws}. The QPM is constructed to describe the equation of state of the QGP obtained from lQCD simulations. In quasi-particle models, the interacting quarks and gluons are effectively treated as an ideal gas of massive, non-interacting quasi-particles. The effective masses of these particles can be interpreted as arising from their interactions. The model was later refined by introducing a temperature-dependent vacuum energy term, $B(T)$, commonly referred to as the bag constant, which restores thermodynamic consistency~\cite{Gorenstein:1995}.\\

The effective temperature- and baryon chemical potential-dependent masses of these quasi-particles are then given by a perturbative approach~\cite{Plumari:2011mk},
\begin{equation}
\begin{alignedat}{2}
& m_g^2(T,\mu_B)  &&= \frac{1}{6} g^2 
  \left[\left(N_c + \tfrac{1}{2} N_f\right) T^2 
  + \frac{N_c}{2\pi^2} \sum_i \mu_i^2\right],\\
& m_{u/d}^2(T,\mu_B) &&= \frac{N_c^2 - 1}{8 N_c}\, g^2
  \left[T^2 + \frac{\mu_{u/d}^2}{\pi^2}\right],\\
& m_s^2(T,\mu_B)     &&= m_{0s}^2 + \frac{N_c^2 - 1}{8 N_c}\, g^2
  \left[T^2 + \frac{\mu_s^2}{\pi^2}\right].
\end{alignedat}
\label{eq:mass_formulae}
\end{equation}
where $N_f = 3$ is the number of quark flavors, $N_c = 3$ is the number of color charges, $m_{u/d}$, $m_s$, and $m_g$ are the masses of the $u/d$ quarks, $s$ quark and gluons, respectively. In this work, we have assumed the masses of the $u$ and $d$ quarks to be equal, and that the strange quark has a rest mass of $m_{0s} = 93.5$ MeV$/c^2$. In the present framework, these expressions are not imposed as exact mass relations. Instead, they are later used only as reference inputs for mass regularization during network training, while the effective quasi-particle masses are learned independently by the neural networks. The quark chemical potentials are related to the baryon chemical potential through the following equation:
\begin{equation}
\begin{alignedat}{2}
& \mu_i &&= b_i\mu_{\mathrm{B}}
\end{alignedat}
\label{eq:muq}
\end{equation}
with $b_i$ being the $i^{th}$ quark's baryon number.
The coupling constant, $g(T)$, is obtained by fitting the EoS to the data obtained from the Wuppertal-Budapest (WB) lQCD group~\cite{Plumari:2011mk},
\begin{equation}
\begin{alignedat}{2}
& g^2(T) &&= \frac{24 \pi^2}{(11N_c - 2N_f)\ln[\lambda(\frac{T}{T_c} - \frac{T_s}{T_c})]}
\end{alignedat}
\label{eq:gt}
\end{equation}
\noindent
here, we choose $\lambda = 2.6$ and $T_s/T_c = 0.57$ as the fit parameters. The deconfinement transition temperature is taken to be $T_c = 155~\mathrm{MeV}$. This parametrized form reproduces the lQCD results with good precision only for $T \ge 1.1T_{\mathrm{c}}$ and, as a consequence, this work is for a temperature range, $T \ge 1.1T_{\mathrm{c}}$~\cite{Puglisi:2014sha}. One should note that the parametrized coupling $g(T)$ in Eq.~(\ref{eq:gt}) is taken from Ref.~\cite{Plumari:2011mk}, where it was obtained by fitting the quasi-particle model to the lattice QCD equation of state. In the present framework, however, the effective quasi-particle masses are learned independently by the neural networks, while Eq.~(\ref{eq:mass_formulae}) is used only to construct the mass-regularization terms. Consequently, the effective quasi-particle masses predicted by the DLQPM framework are not required to satisfy the perturbative mass relations given in Eq.~(\ref{eq:mass_formulae}).

\subsection{Equation of State}
Within the grand canonical ensemble, the complete partition function for the QGP, comprising deconfined gluons and light quarks ($u$, $d$, $s$), can be written as;
{\small
\begin{equation}
\ln Z(T,\mu_{\mathrm{B}}) = \ln Z_g(T,\mu_{\mathrm{B}}) + \sum_{i=u,d,s} \ln Z_{q(\bar{q})_i}(T,\mu_{\mathrm{B}}).
\label{eq:lnZ}
\end{equation}
}
Here, the first term accounts for the gluonic degrees of freedom, and the sum runs over the three lightest quark flavors and their antiquark counterparts. The separation into individual contributions reflects the quasi-particle picture, in which the strongly interacting medium is described in terms of gluons and quarks carrying medium-dependent effective masses. \\

The individual partition functions for the gluons, quarks, and antiquarks are given by 
{\footnotesize
\begin{flalign}
\ln Z_g(T,\mu_{\mathrm{B}})
&=-\frac{d_gV}{2\pi^2}
\int_0^\infty\! p^2dp \nonumber\\[1pt]
&\quad\times \ln\!\left[1-\exp\!\left(
-\frac{1}{T}\sqrt{p^2+m_g^2(T,\mu_{\mathrm{B}})}
\right)\right]
&&\label{eq:Zg}
\end{flalign}

\begin{flalign}
\ln Z_{q_i}(T,\mu_{\mathrm{B}})
&=+\frac{d_{q_i}V}{2\pi^2}
\int_0^\infty\! p^2dp \nonumber\\[1pt]
&\quad\times \ln\!\left[1+\exp\!\left(
-\frac{1}{T}\!\left(
\sqrt{p^2+m_{q_i}^2(T,\mu_{\mathrm{B}})}
-\mu_{q_i}
\right)
\right)\right]
&&\label{eq:Zq}
\end{flalign}

\begin{flalign}
\ln Z_{\bar{q}_i}(T,\mu_{\mathrm{B}})
&=+\frac{d_{\bar{q}_i}V}{2\pi^2}
\int_0^\infty\! p^2dp \nonumber\\[1pt]
&\quad\times \ln\!\left[1+\exp\!\left(
-\frac{1}{T}\!\left(
\sqrt{p^2+m_{\bar{q}_{i}}^2(T,\mu_{\mathrm{B}})}
+ \mu_{\bar{q}_i}
\right)
\right)\right]
&&\label{eq:Zqbar}
\end{flalign}
}

\noindent
corresponding to the equations,  $d_g$ and $d_{q_{i}} $ are the gluon and quark-antiquark degeneracy factors, and $V$ is the volume of the system. The $m_g(T,\mu_{\mathrm{B}})$ and $m_{q_{i}}(T,\mu_{\mathrm{B}})$ are the medium-dependent thermal masses of the gluons and quark-antiquark pairs. The thermal mass $m_g(T,\mu_{\mathrm{B}})$ encapsulates the non-perturbative interactions of gluons with the surrounding medium, replacing the bare massless gluon of perturbative QCD with an effective excitation that carries the collective properties of the QGP.\\

Further, the thermodynamic variables for the deconfined hot QCD matter can be obtained using the partition function, as defined below:
{\small
\begin{equation}
P(T,\mu_{\mathrm{B}}) = T \left( \frac{\partial \ln{Z(T,\mu_{\mathrm{B}})}}{\partial V}\right)_{T,\mu_{\mathrm{B}}}
\label{P}
\end{equation}

\begin{equation}
n_{\mathrm{B}}(T,\mu_{\mathrm{B}}) = \left(\frac{\partial P(T,\mu_{\mathrm{B}})}{\partial \mu_{\mathrm{B}}}\right)_T
\label{nb}
\end{equation}

\begin{equation}
s(T,\mu_{\mathrm{B}}) = \left(\frac{\partial P(T,\mu_{\mathrm{B}})}{\partial T}\right)_{\mu_{\mathrm{B}}}
\label{s}
\end{equation}

\begin{equation}
\epsilon(T,\mu_{\mathrm{B}}) = Ts(T,\mu_{\mathrm{B}}) - P(T,\mu_{\mathrm{B}}) + \mu_{\mathrm{B}}\, n_{\mathrm{B}}
\label{e}
\end{equation}

\begin{equation}
\Delta(T,\mu_{\mathrm{B}}) = \epsilon(T,\mu_{\mathrm{B}}) - 3P(T,\mu_{\mathrm{B}})
\label{trace}
\end{equation}
}
where $P$, $n_{\rm B}$, $s$, $\epsilon$, and $\Delta$ are pressure, baryon number density, entropy density, energy density, and trace anomaly, respectively. Now, using the above-mentioned thermodynamic relations
the speed of sound squared $\left( \cs \right)$ and the specific heat $\left( \Cv \right)$ are obtained as:

{\small
\begin{equation}
\cs = \left(\frac{\partial P}{\partial \epsilon}\right)_{s/n_{\mathrm{B}}}
\label{eq:cs2}
\end{equation}
\begin{equation}
\Cv = \left(\frac{\partial \epsilon}{\partial T}\right)_V
\label{eq:Cv}
\end{equation}
}

\subsection{Transport Properties}
The space-time evolution of the QGP fireball created in ultrarelativistic heavy-ion collisions is governed by two key ingredients: the initial conditions of the collision and the transport properties of the deconfined medium. Among the most phenomenologically relevant transport coefficients are the shear viscosity ($\eta$) and bulk viscosity ($\zeta$), which regulate the momentum isotropization and pressure equilibration of the expanding medium, respectively. As demonstrated in Refs.~\cite{Schenke:2011bn, Gardim:2020mmy}, the anisotropic flow coefficients $v_n$ serve as direct experimental observables that encode the geometry of the initial state and the subsequent hydrodynamic response, and are highly sensitive to both $\eta$ and $\zeta$. In particular, a finite shear viscosity affects the anisotropic flow, while bulk viscosity modifies the longitudinal expansion of the fireball. The cooling rate of the QGP is likewise sensitive to these viscosities, as larger dissipative corrections slow down the hydrodynamic evolution that drives the collective expansion~\cite{Sahoo:2023xnu}. Beyond the viscous phenomena, the transport of conserved charges and heat within the QGP plays an equally significant role in the medium dynamics. The electrical conductivity ($\sigma_{\mathrm{el}}$) governs the response of the deconfined medium to the intense, albeit transient, electromagnetic fields generated in the early stages of a heavy-ion collision. It controls the magnitude and lifetime of the induced electric field. It restrains the diffusion of electric charge through the medium, with direct implications for the generation and sustainability of the electromagnetic field~\cite{Singh:2024emy}. The thermal conductivity ($\kappa_T$), on the other hand, controls the flow of heat across the medium. Although often overlooked in hydrodynamic modeling, $\kappa_T$ has been shown to indirectly influence collective flow observables, including the elliptic flow coefficient $v_2$~\cite{Singh:2023ues}, by modifying the local temperature profiles during the hydrodynamic evolution.\ Together, these transport coefficients encode the microscopic interaction structure of the QGP and are indispensable inputs for constructing a quantitative description of the medium evolution from the initial thermalization to the final hadronization. The evaluation of these viscous phenomena is further contextualized by recent comprehensive advancements in the fundamental relativistic theory of fluid viscosity~\cite{Zaccone:2024zel}.\\

 Now, to compute the viscous transport coefficients, we employ the relativistic Boltzmann transport equation within the relaxation-time approximation~(RTA)~\cite{Plumari:2011mk, Sasaki:2008fg}, which provides a tractable yet physically motivated framework for estimating $\eta$, $\zeta$, $\sigma_{\mathrm{el}}$, and $\kappa_T$ in terms of the quasi-particle thermal masses and their medium-dependent relaxation times.

{\small
\begin{equation}
\begin{aligned}
\eta = \frac{1}{15T} \sum_i d_i \int \frac{d^3p}{(2\pi)^3} \tau_i \frac{p^4}{E_i^2} f_i(1 \mp f_i)
\end{aligned}
\label{eq:eta}
\end{equation}

\begin{equation}
\begin{aligned}
\zeta &= -\frac{1}{3T}
\sum_i d_i
\int \frac{d^3p}{(2\pi)^3}
\Bigg\{
\frac{m_i^2}{E_i}
\, \tau_i f_i(1 \mp f_i)
\\
&\quad \times
\Bigg[
\frac{p^2}{3E_i}
- \left(\frac{\partial P}{\partial \epsilon}\right)_{n_B}
\left(
E_i
- T \frac{\partial E_i}{\partial T}
- \mu_{\mathrm{B}} \frac{\partial E_i}{\partial \mu_{\mathrm{B}}}
\right)
\\
&\quad
+ \left(\frac{\partial P}{\partial n_B}\right)_{\epsilon}
\frac{\partial E_i}{\partial \mu_{\mathrm{B}}}
\Bigg] - \frac{m_i^2}{E_i}
\, \tau_i f_i(1 \mp f_i)
\left(\frac{\partial P}{\partial n_B}\right)_{\epsilon}
\Bigg\}
\end{aligned}
\label{eq:xi}
\end{equation}
}

Here, $f_i$ is the Fermi-Dirac and Bose-Einstein distribution functions for quarks and gluons, respectively. 
The $\tau_{i}$ denotes the relaxation time for the $i^{th}$ species. Moreover, the electrical and thermal conductivity are given by the following relations~\cite{Madni:2024ubw, Singh:2023pwf, Abhishek:2020wjm},
{\small
\begin{equation}
\sigma_{el} = \frac{1}{3T} 
\sum_i d_i 
\int \frac{d^3p}{(2\pi)^3} \,
\tau_i q_i^2\frac{p^2}{E_i^2} f_i(1 - f_i)
\label{eq:conductivity}
\end{equation}

\begin{equation}
\begin{aligned}
\kappa_T = \frac{1}{3T^2} 
\sum_i d_i 
\int \frac{d^3p}{(2\pi)^3} \,
\tau_i \frac{p^2}{E_i^2} (E_i-b_ih)^2f_i(1 \mp f_i)
\label{eq:conductivity}
\end{aligned}
\end{equation}
}\par
where $q_{i}$ is the electric charge of the $i^{th}$ parton, and enthalpy per particle is given by $h=\frac{\epsilon + P}{n_{\rm B}}$. For this study, we use a parametrized form of relaxation time given as~\cite{Plumari:2011mk},

\begin{equation}
\begin{aligned}
\tau_q^{-1}
&= \frac{N_c^2-1}{N_c} \frac{g^2T}{8\pi} \ln{\frac{2k}{g^2}}, \ \ \ \
\tau_g^{-1} = 2 N_c\frac{g^2T}{8\pi}\ln{\frac{2k}{g^2}}
\end{aligned}
\label{Eq:tau}
\end{equation}

We have chosen the $k=18.5$ value in such a way that Eq.~(\ref{Eq:tau}) yields a minimum value of $\eta/s$ of $1/4\pi$ satisfying the KSS bound. 

The RTA is widely adopted in transport calculations of this kind for its simplicity and computational tractability, and it is the approach we follow here within the DLQPM framework. We note that alternative treatments, such as the Chapman–Enskog expansion and the Green–Kubo formalism, have been shown to give more accurate estimates for the shear viscosity of a one-component fluid~\cite{Plumari:2012ep}, and that for a multi-component system such as the QGP these differences can be more pronounced, especially for the bulk viscosity, since the transport coefficients of a mixture are not simply additive in the single-component contributions~\cite{Parisi:2025gwq,Parisi:2026ufk}. A quantitative comparison along these lines would be a natural extension of the present work.

\section{Methodology}
\label{Methodology}

The medium-dependent quasi-particle masses are extracted using residual neural networks (ResNets) trained to reproduce the lattice QCD equation of state~\cite{Borsanyi:2021sxv}. The computational framework for this consists of three independent ResNets, each dedicated to learning the thermal mass of a distinct quasi-particle species: gluons, light quarks ($u$ and $d$), and strange quarks ($s$), following the architecture introduced in Refs.~\cite{Li:2022ozl, Li:2025csc, Jamal:2025gjy}. Each network receives a two-dimensional input vector $(T, \mu_{\mathrm{B}})$, representing the temperature and baryon chemical potential of the medium, and outputs a single positive scalar corresponding to the effective thermal mass of the respective quasi-particle species, namely $m_g(T,\mu_{\mathrm{B}})$, $m_{u/d}(T,\mu_{\mathrm{B}})$, and $m_s(T,\mu_{\mathrm{B}})$. The positivity of the output is enforced by construction, ensuring the physical consistency of the learned masses across the entire $(T, \mu_{\mathrm{B}})$ parameter space.\\

 Further, the learned quasi-particle masses are subsequently used as inputs to the grand canonical partition function, as defined in  Eqs.~(\ref{eq:lnZ})--(\ref{eq:Zqbar}), from which thermodynamic 
observables of the QGP are derived. Specifically, we compute the pressure  $P$, energy density $\epsilon$, entropy density $s$, and trace anomaly  $\Delta = \epsilon - 3P$, which serves as a sensitive probe of the non-perturbative structure of the QCD medium near the deconfinement transition. The resulting thermodynamic quantities are benchmarked against the lQCD results of Ref.~\cite{Borsanyi:2021sxv},  providing a stringent test of the network's ability to generalize beyond the training data and accurately capture the thermodynamic structure of the QGP at finite baryon chemical potential.\\

\begin{figure*}
    \centering
    \includegraphics[width=.9\linewidth]{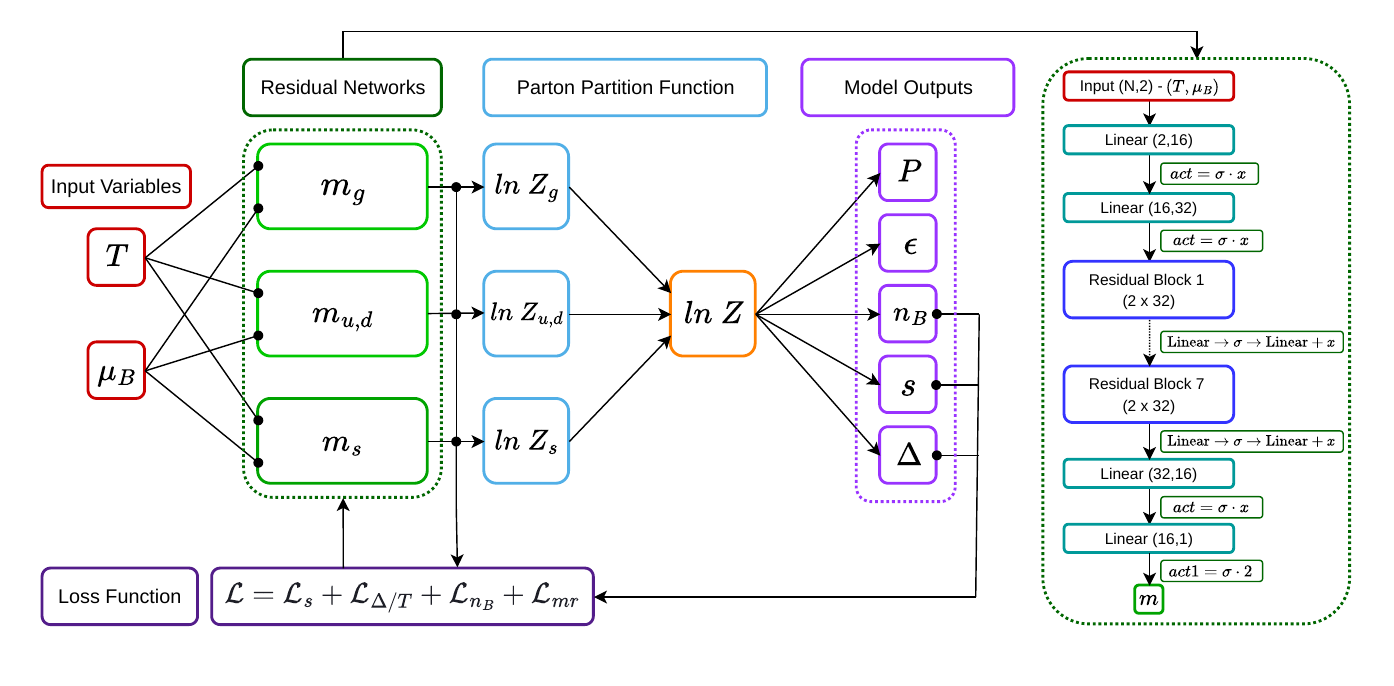}
    \caption{Schematic diagram of the deep-learning-assisted quasi-particle framework. The inputs $(T,\mu_B)$ are processed by three residual neural networks to predict the effective quasi-particle masses, which are used to compute thermodynamic observables through the partition function. The predicted quantities are used to construct the loss function, whose gradients are back-propagated to update the network parameters during training.}
    \label{Fig:Model_arch}
\end{figure*}

The architecture and training workflow of the model are illustrated in Fig.~\ref{Fig:Model_arch}. The three ResNets are initialized with random weight parameters and trained simultaneously in a supervised learning setting. The deviation between the predicted observables and the lQCD benchmark values~\cite{Borsanyi:2021sxv} is quantified via a total loss function $\mathcal{L}$, which is minimized through backpropagation using a gradient-based optimizer.\ Training continues until convergence, yielding an effective mapping $(T, \mu_{\mathrm{B}}) \mapsto \{m_g, m_{u/d}, m_s\}$ that is thermodynamically consistent with the lQCD equation of state. The total loss function is defined as,
\begin{align*}
\mathcal{L} = \mathcal{L}_{s} + \mathcal{L}_{\Delta/T} + \mathcal{L}_{n_B}  + \mathcal{L}_{mr}
\end{align*}
where $\mathcal{L}_{s}$, $\mathcal{L}_{\Delta/T}$, and $\mathcal{L}_{n_B}$ are the loss functions, given as,
\begin{equation}
\begin{aligned}
  \mathcal{L}_{s}  &= \frac{1}{N} \sum_i \left| {(s)}_{\text{pred}} - {(s)}_{\text{WB}} \right|_i \\
  \mathcal{L}_{\Delta/T} &= \frac{1}{N} \sum_i \left| \left(\frac{\Delta}{T}\right)_\text{pred} - \left(\frac{\Delta}{T}\right)_\text{WB}\right|_{i} \\
  \mathcal{L}_{n_B}  &= \frac{1}{N} \sum_i \left| {(n_B)}_{\text{pred}} - {(n_B)}_{\text{WB}} \right|_i
\end{aligned}
\end{equation}

\begin{table}[h]
\centering
\caption{Training hyper-parameters used to optimize the residual neural networks.}
\label{tab:training_hyperparameters}
\begin{ruledtabular}
\begin{tabular}{l c}
\textbf{Hyperparameter} & \textbf{Value} \\
\hline
Optimizer & AdamW \\
Learning rate & $0.001$ \\
Learning rate scheduler & StepLR \\
Scheduler step size & $2000$ epochs \\
Scheduler decay factor $\gamma$ & $0.92$ \\
Number of training epochs & $5000$ \\
Loss function & Mean Absolute Error (L1) \\
Train/validation split & $80\% / 20\%$ \\
\end{tabular}
\end{ruledtabular}
\end{table}
However, constraining the quasi-particle masses solely through the loss function constructed from thermodynamic observables leads to non-unique solutions for $m_{g}$, $m_{u/d}$, and $m_{s}$. This non-uniqueness arises because multiple combinations of these masses can reproduce the same set of thermodynamic observables. To address this non-uniqueness, we introduce additional mass-regularization terms in the total loss function~($\mathcal{L}_{mr}$). These terms enforce further constraints on the relative mass hierarchy, guided by QPM results~\cite{Plumari:2011mk}. In this way, the networks are constrained to solutions that remain consistent with the established quasi-particle mass hierarchy while still accurately reproducing thermodynamic variables obtained from the lattice. The mass regularization loss is defined as,
\begin{small}
\begin{equation}
\begin{aligned}
\mathcal{L}_{mr} &= R_{g/ud} + R_{g/s}, \\[2mm]
R_{g/ud} &= \beta \left| 
    \left( \frac{m_g}{m_{u/d}} \right)_{\mathrm{pred}}
    - 
    \left( \frac{m_g}{m_{u/d}} \right)_{\mathrm{QPM}}
\right|, \\[1mm]
R_{g/s} &= \beta \left| 
    \left( \frac{m_g}{m_s} \right)_{\mathrm{pred}}
    - 
    \left( \frac{m_g}{m_s} \right)_{\mathrm{QPM}}
\right|.
\label{Eq:mass_reg}
\end{aligned}
\end{equation}
\end{small}

where $R_{g/ud}$ and $R_{g/s}$ are the mass ratios computed via Eq.~(\ref{eq:mass_formulae}), and $\beta = 0.01$ controls the relative strength of the regularization terms. The required derivatives are then computed via automatic differentiation using the AutoGrad engine~\cite{Paszke:2019xhz,PyTorch:Autograd}, while the momentum integrals in the partition function are evaluated using 50-point Gaussian quadrature. The networks are trained for $5000$ epochs using the mean absolute error (MAE) loss with an $80\%/20\%$ train-validation split. The AdamW optimizer~\cite{Kingma:2014vow,Loshchilov:2017bsp,PyTorch:AdamW}  is employed with an initial learning rate of $\alpha = 0.001$, combined with a StepLR scheduler~\cite{PyTorch:StepLR} that decays 
the learning rate by a factor of $\gamma = 0.92$ every $2000$ epochs, ensuring stable convergence near the loss minimum. The complete set of training hyperparameters is summarized in Table~\ref{tab:training_hyperparameters}, and further details regarding the neural network architecture and training procedure are provided in the Appendix~\ref{appendix}.

The trained model is subsequently used to extract the  $(T, \mu_{\mathrm{B}})$-dependent quasi-particle mass evolution and 
to compute the thermodynamic and transport properties of the QGP across the relevant parameter space.

\begin{figure*}
    \includegraphics[width=0.32\linewidth]{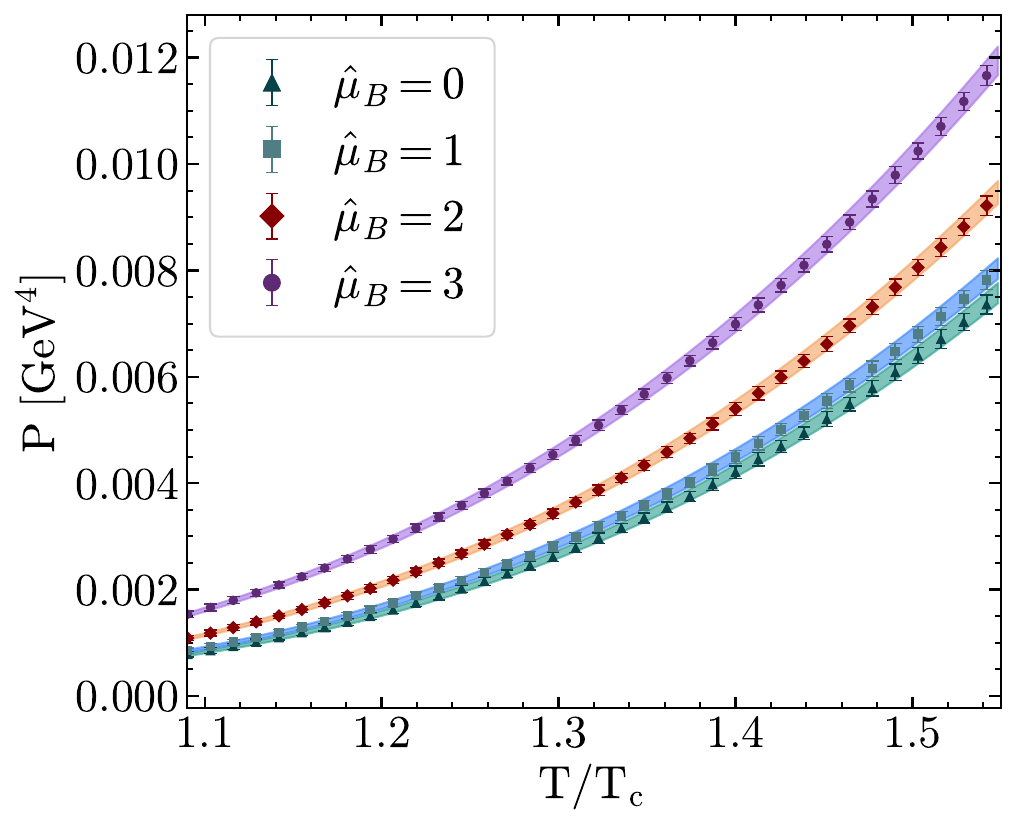}
    \includegraphics[width=0.32\linewidth]{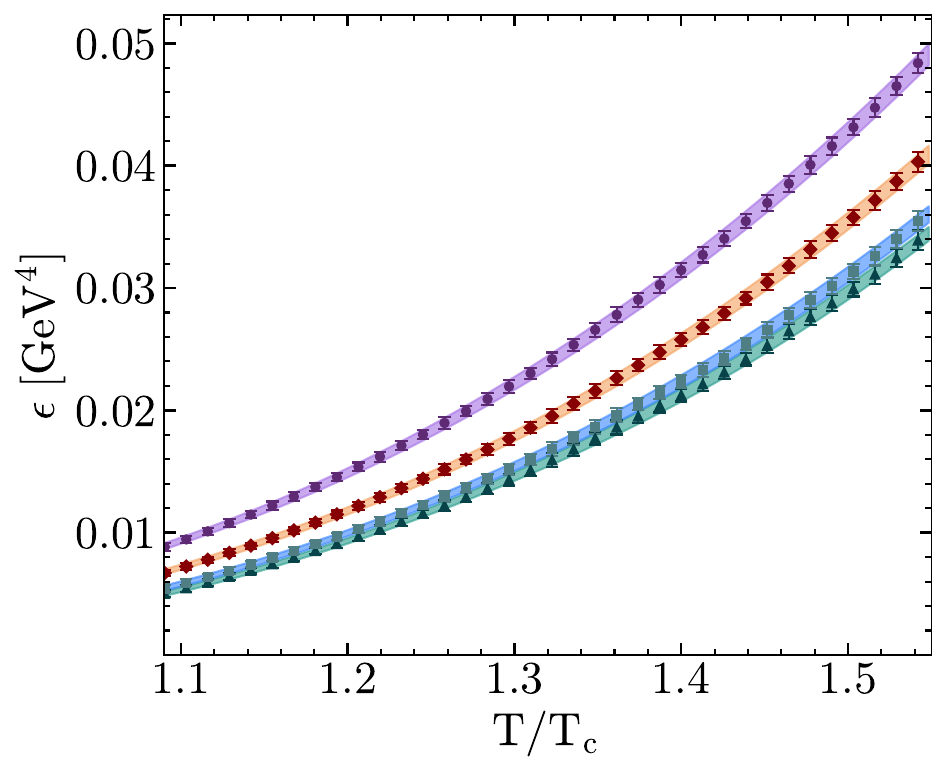}
    \includegraphics[width=0.32\linewidth]{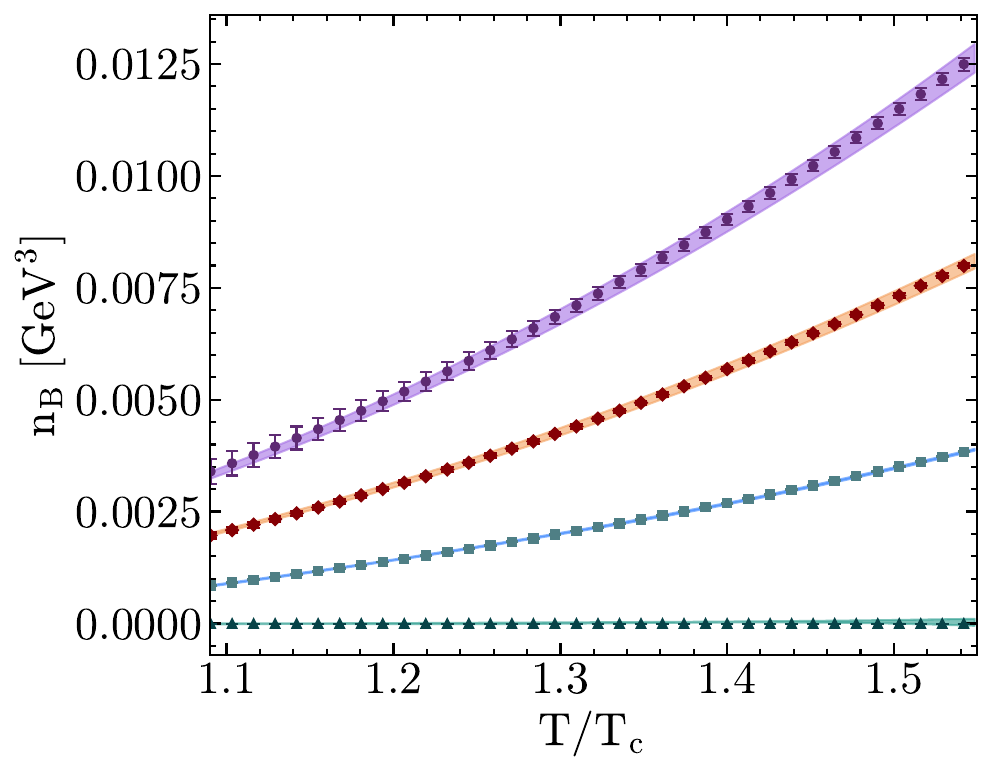}\\[4pt]
    \includegraphics[width=0.32\linewidth]{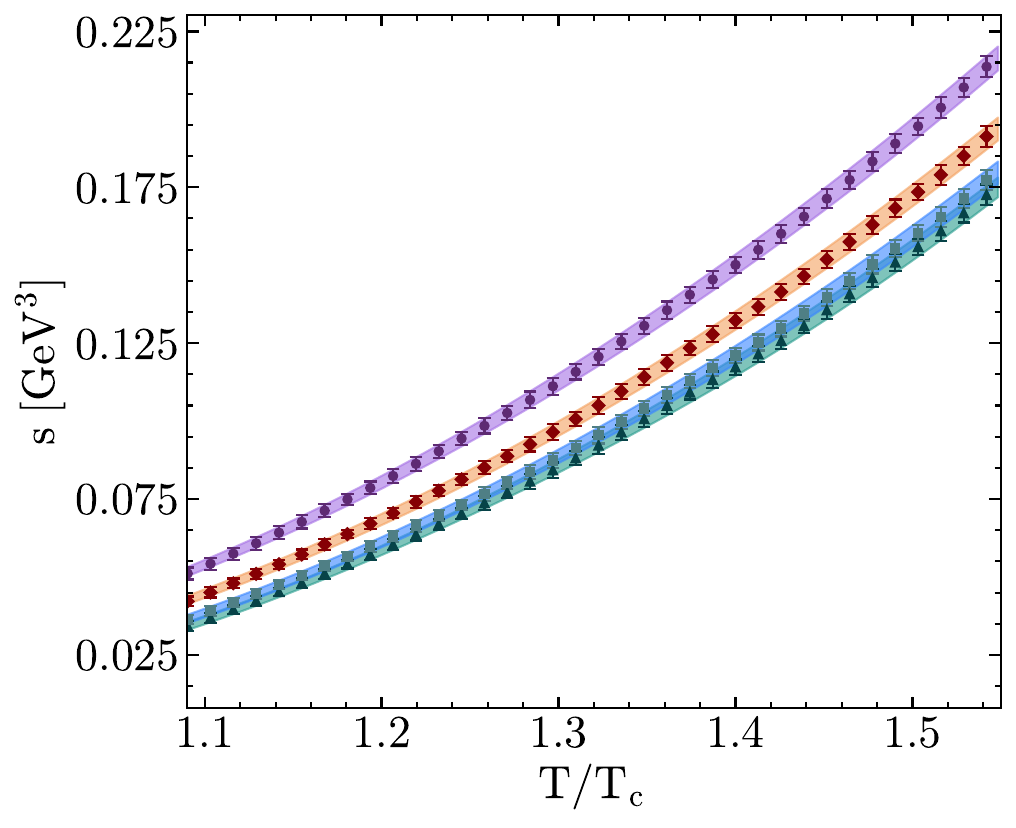}
    \includegraphics[width=0.32\linewidth]{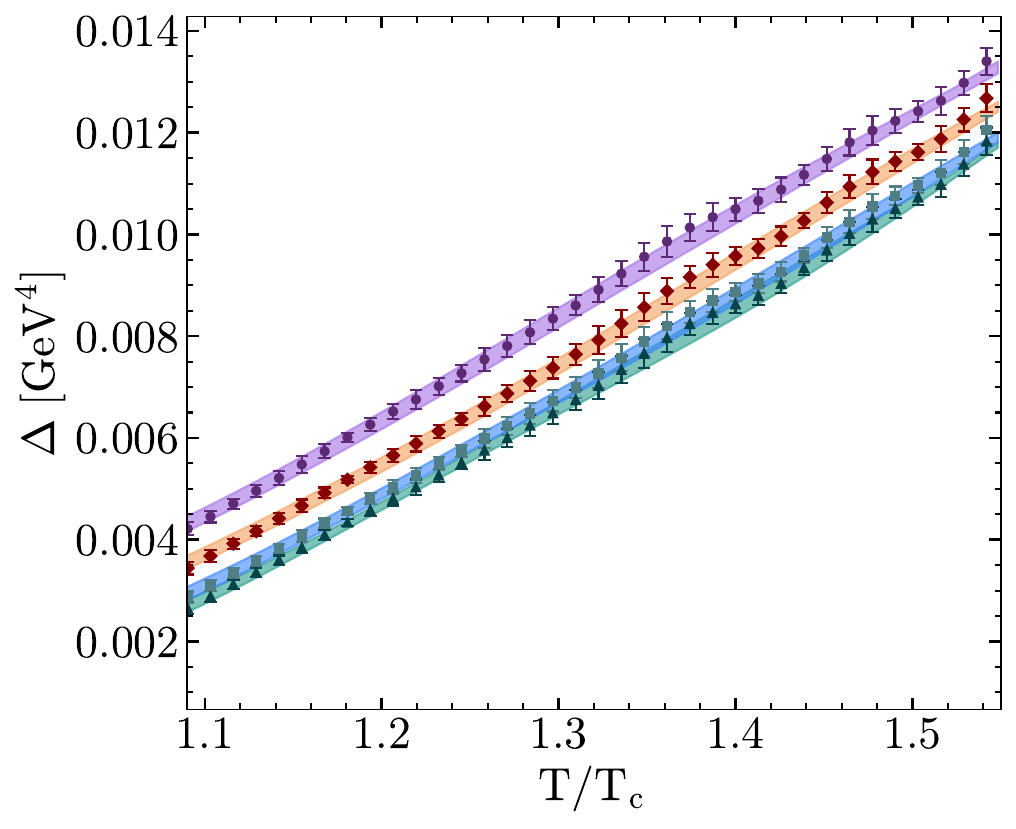}
    \caption{Pressure $(P)$, energy density $(\epsilon)$, baryon number density $(n_B)$, entropy density $(s)$, and trace anomaly $(\Delta)$ as a function of temperature at at different values of $\mucapB=\mu_{\rm B}/T$. The colored bands represent the model predictions, while the markers denote the Wuppertal-Budapest lQCD results~\cite{Borsanyi:2021sxv}.}
    \label{Fig:Thermo}
\end{figure*}
\begin{figure*}[t]
    \centering
    \includegraphics[width=0.32\linewidth]{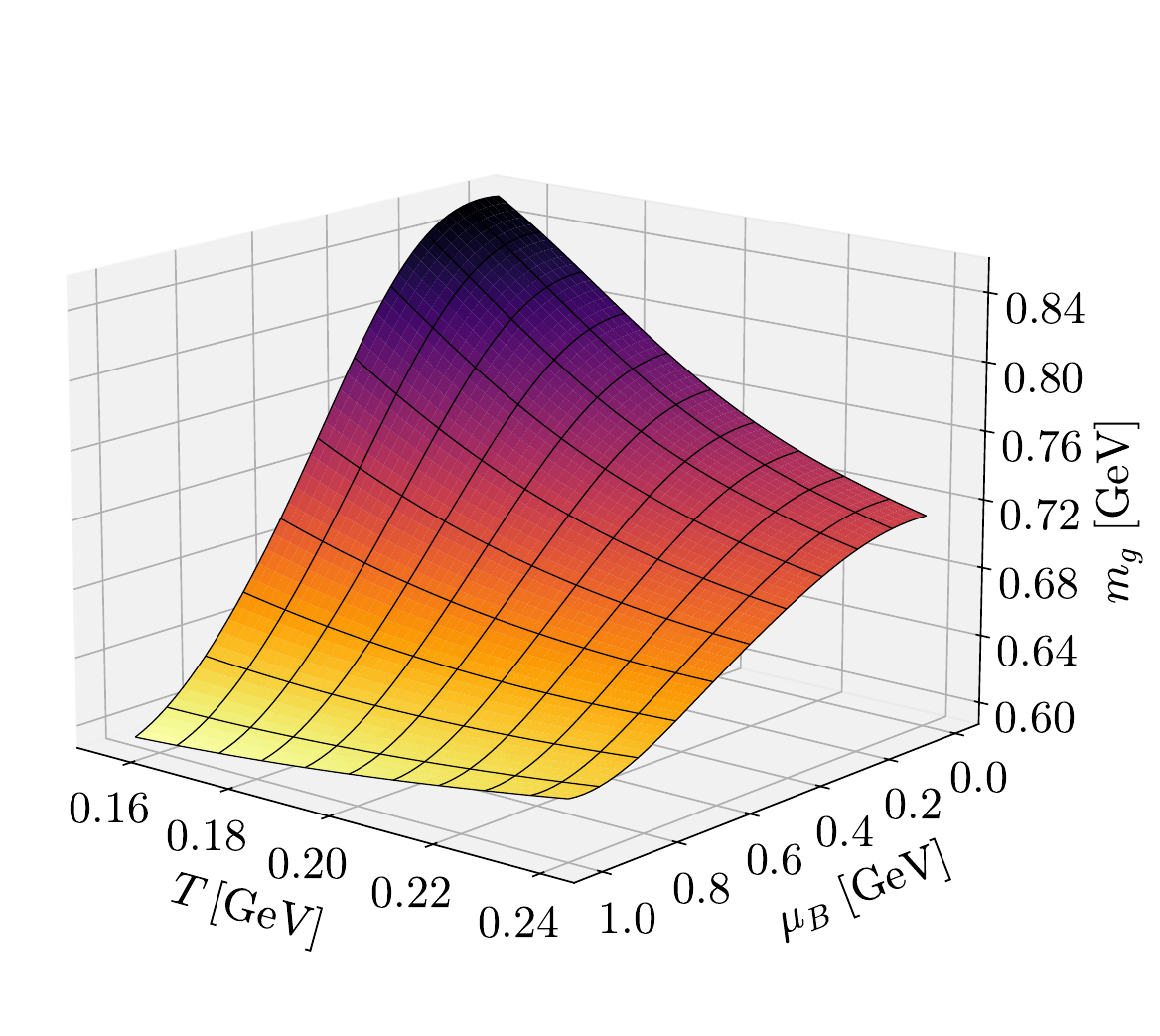}
    \includegraphics[width=0.32\linewidth]{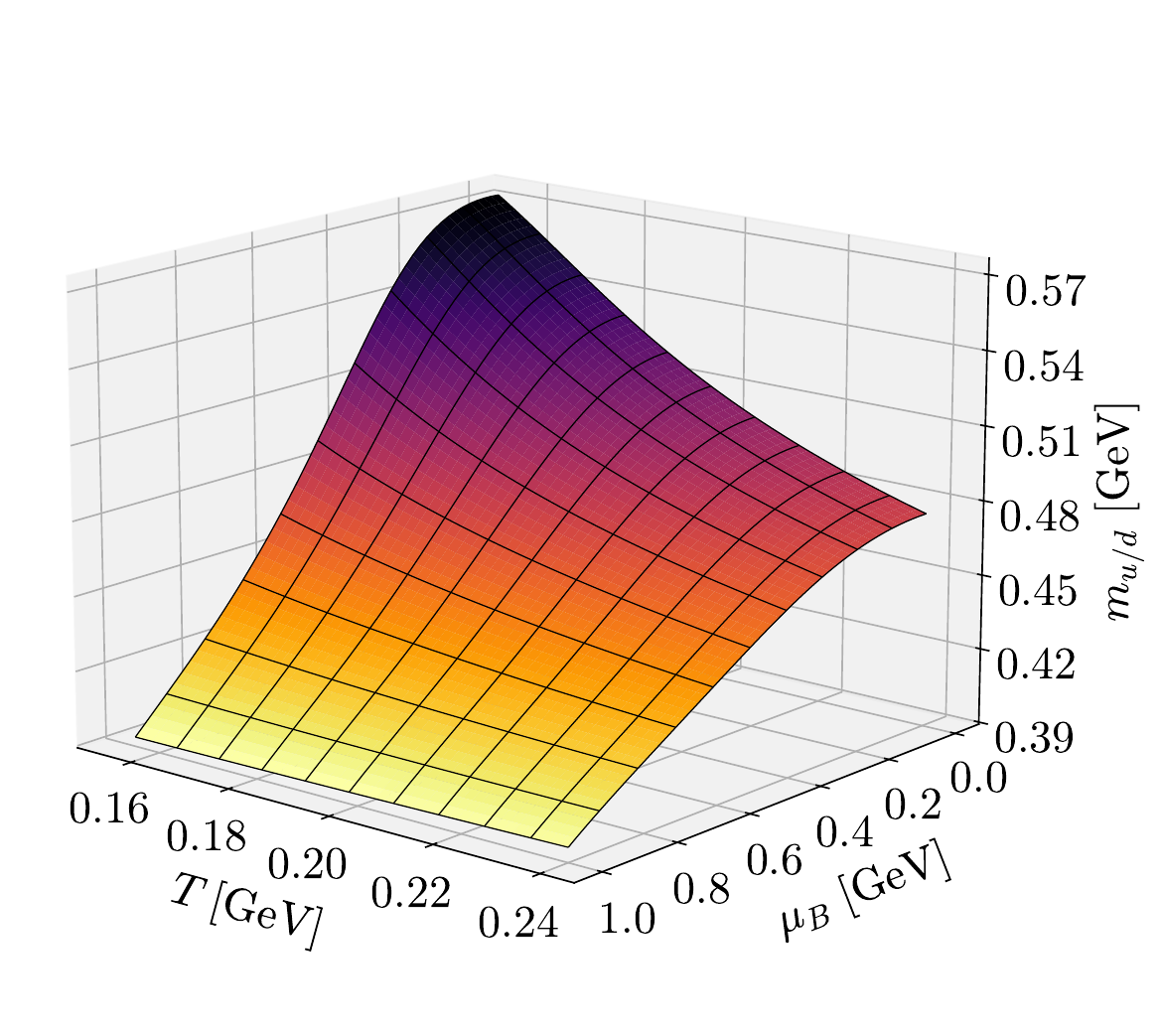}
    \includegraphics[width=0.32\linewidth]{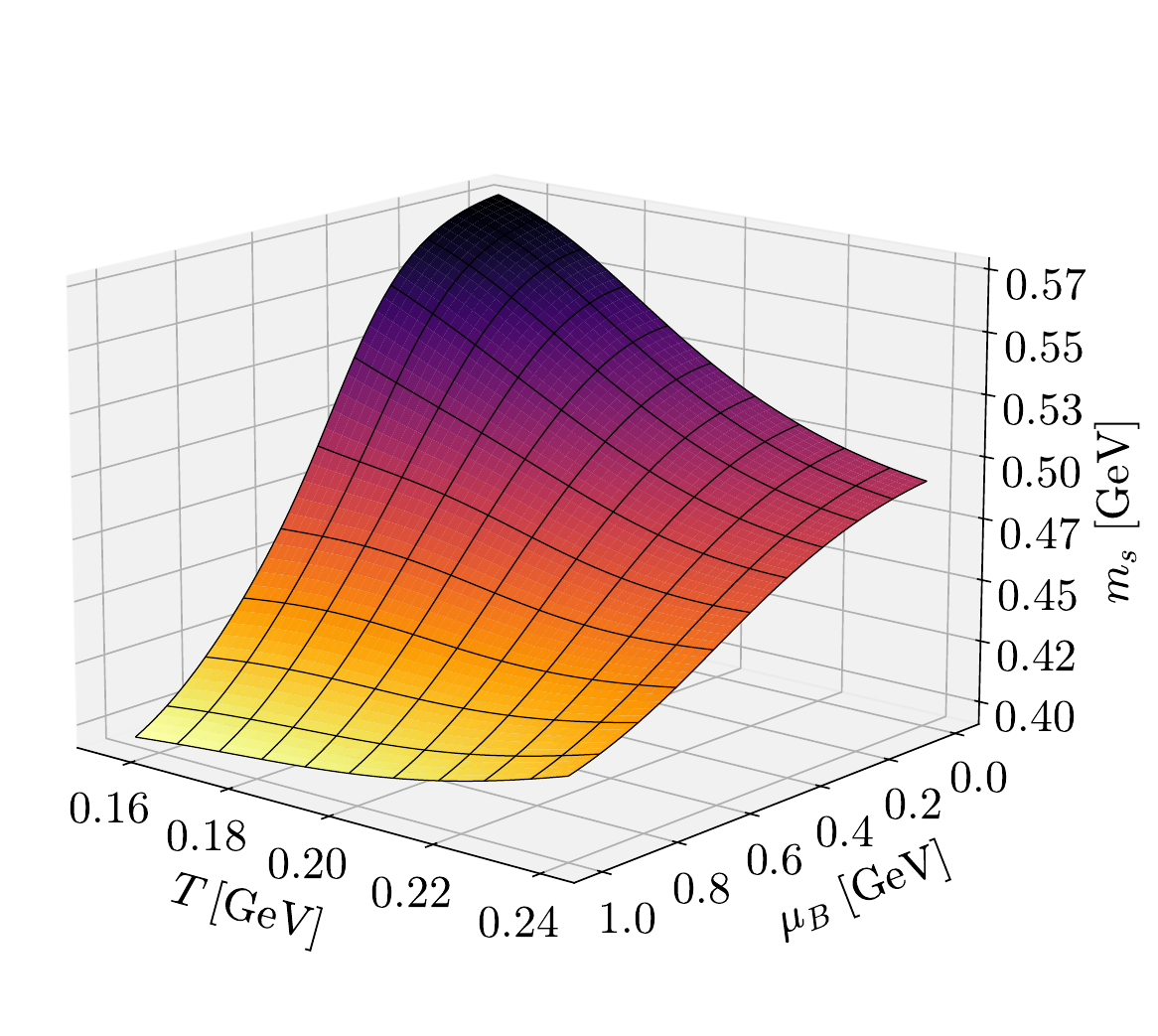}
    \caption{Three-dimensional surface plots of the effective quasi-particle masses predicted by the DNN as functions of temperature and baryon chemical potential. The left panel corresponds to the gluon mass $m_g$, the middle panel represents the light quark mass $m_{u/d}$, and the right panel shows the strange quark mass $m_s$.}
    \label{Fig:3DMass}
\end{figure*}
\section{Results and Discussions}
\label{Results}

Using the trained quasi-particle masses obtained from the residual networks, we evaluate the partition function and compute the corresponding thermodynamic observables and transport properties across the $(T, \mu_B)$ plane for different values of $\hat{\mu}_{\rm B}$ ($=\mu_{\rm B}/T$). The results for all the thermodynamic and transport properties are plotted as bands. The upper and lower bounds on these bands represent the uncertainties on the training data in lQCD. Fig.~\ref{Fig:Thermo} shows the thermodynamic observables $P$, $\epsilon$, $n_B$, $s$, and $\Delta$ as functions of scaled temperature $(T/T_{c})$, where $T_{c} = 155$ MeV. It is important to note that the model is trained using thermodynamic observables derived from the partition function. The loss function is constructed using the entropy density $s$, baryon number density $n_B$, and the trace anomaly $\Delta = \epsilon - 3P$, which indirectly constrain the pressure and energy density. We observe a very good agreement between the estimated thermodynamic observables and lQCD calculations~\cite{Borsanyi:2021sxv}. Moreover, a clear dependence on $\hat{\mu}_{\rm B}$ can be observed throughout the entire temperature range for all the thermodynamic observables. \\
\begin{figure}
    \centering
    \includegraphics[width=0.45\textwidth]{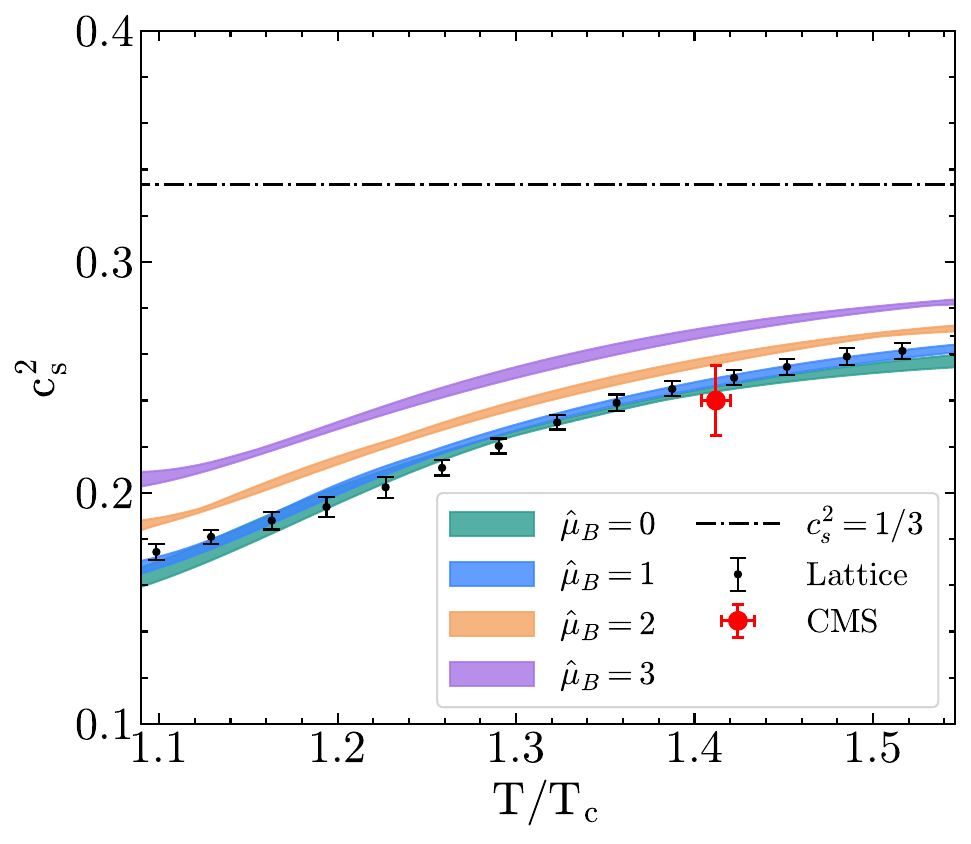}
    \includegraphics[width=0.45\textwidth]{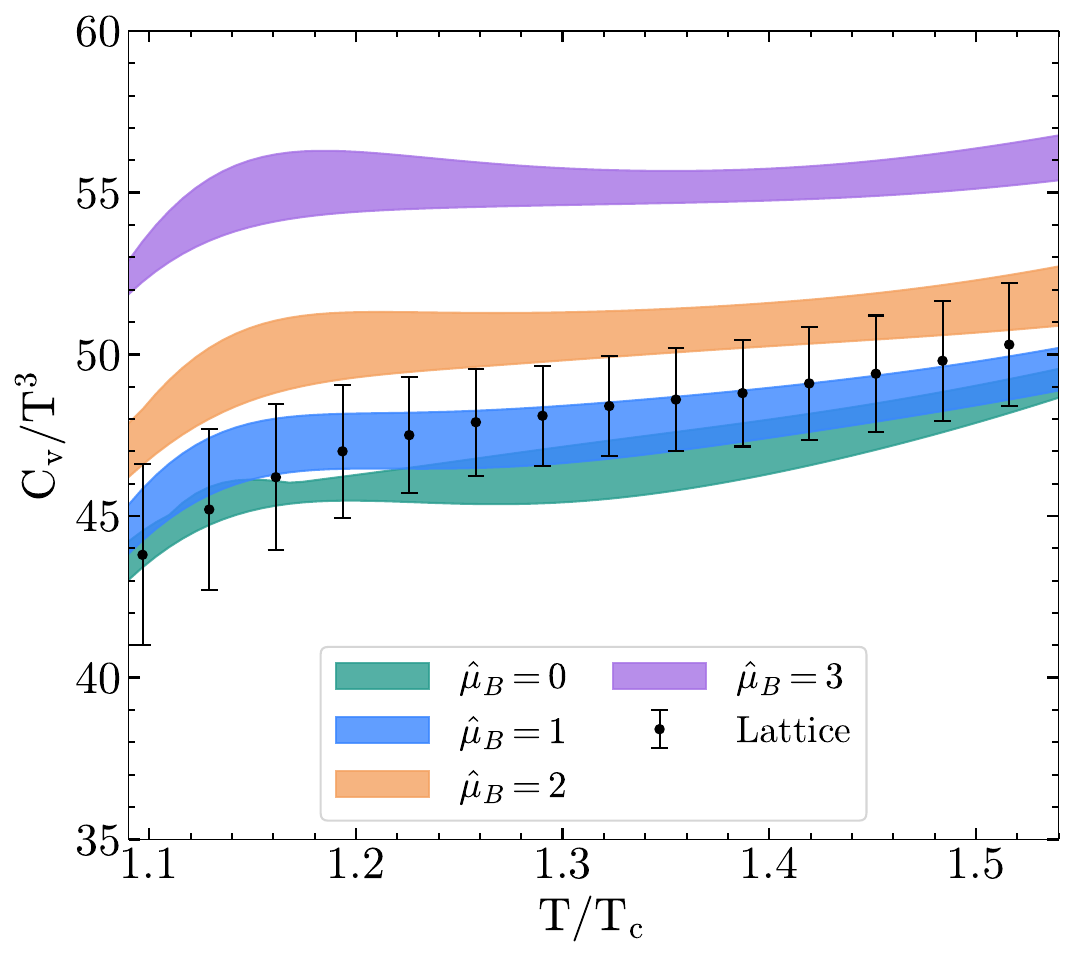}
    \caption{(Top) Speed of sound squared $\cs$ and (bottom) scaled specific heat $\Cv/T^3$ as functions of $\TbyTc$ from the DNN-based quasi-particle model at different $\mucapB$ values. In the top panel, lQCD results from Ref.~\cite{Borsanyi:2013bia} and the CMS measurement~\cite{CMS:2024sgx} are shown for comparison, while the dashed line indicates the Stefan--Boltzmann limit. In the bottom panel, lQCD results from Ref.~\cite{HotQCD:2014kol} are shown.}
    \label{Fig:cs2cvt3}
\end{figure}

Fig.~\ref{Fig:3DMass} depicts the 3D surface maps of the effective thermal masses of the quasi-particles as functions of temperature and baryon chemical potential. The smooth variation of the extracted masses over the entire $(T,\muB)$ plane indicates that the residual networks have learned a stable and physically consistent mapping without oscillations or discontinuities.  All three masses, $\mg$, $\mud$, and $\ms$, exhibit qualitatively similar behavior across the $T-\muB$ plane. At $\muB = 0$, a decreasing trend with temperature is observed for all the masses. In contrast, at higher $\muB$, the temperature dependence becomes comparatively weaker. In addition, a very strong dependence on $\muB$ is observed for all temperature values, particularly in the lower temperature range around $T\sim160$ MeV. The stronger $\muB$ dependence at lower temperatures indicates that the medium interactions, represented by the effective quasi-particle masses, are more sensitive to the baryon chemical potential near the deconfinement temperature. Moreover, the hierarchical ordering $\mg > \ms > \mud$ is preserved throughout the entire phase space, which demonstrates that the imposed mass regularization successfully constrains the effective quasi-particle masses.\\
\begin{figure}
    \centering
    \includegraphics[width=0.45\textwidth]{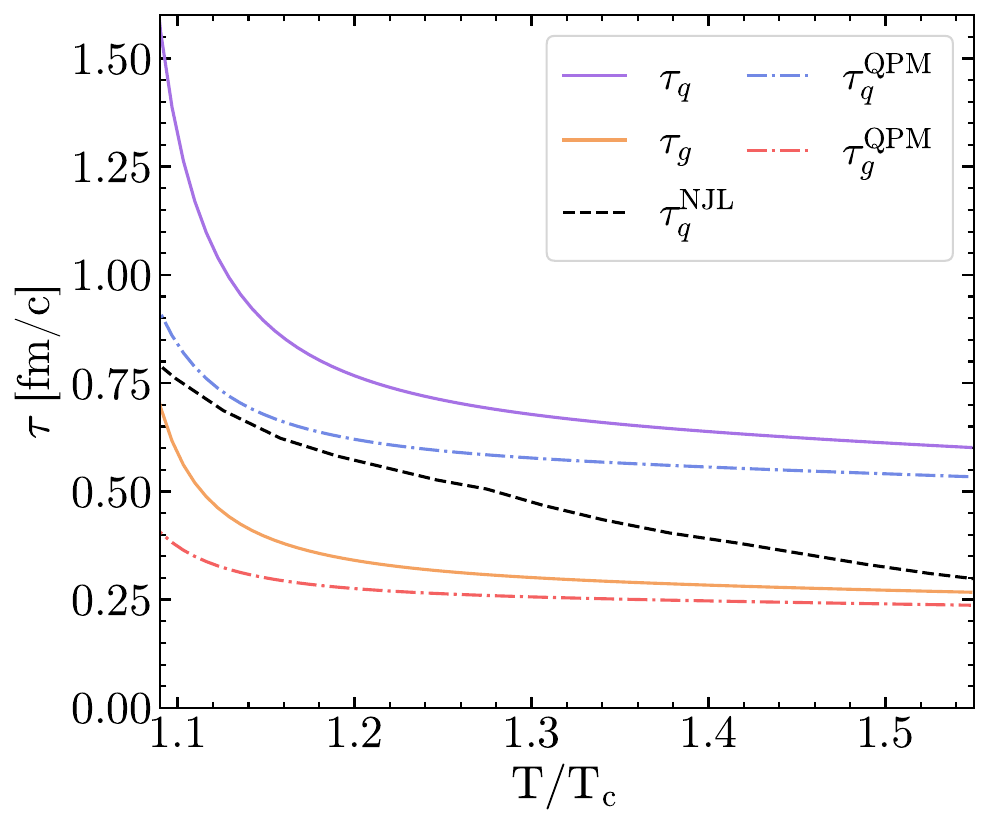}
    \caption{The relaxation times $\tau_q$ for quarks and $\tau_g$ for gluons as functions of $\TbyTc$. The results are compared with the relaxation times obtained in the QPM~\cite{Plumari:2011mk} and the NJL model~\cite{Marty:2013ita}.}
    \label{Fig:tau}
\end{figure}

\begin{figure}
    \centering
    \includegraphics[width=0.45\textwidth]{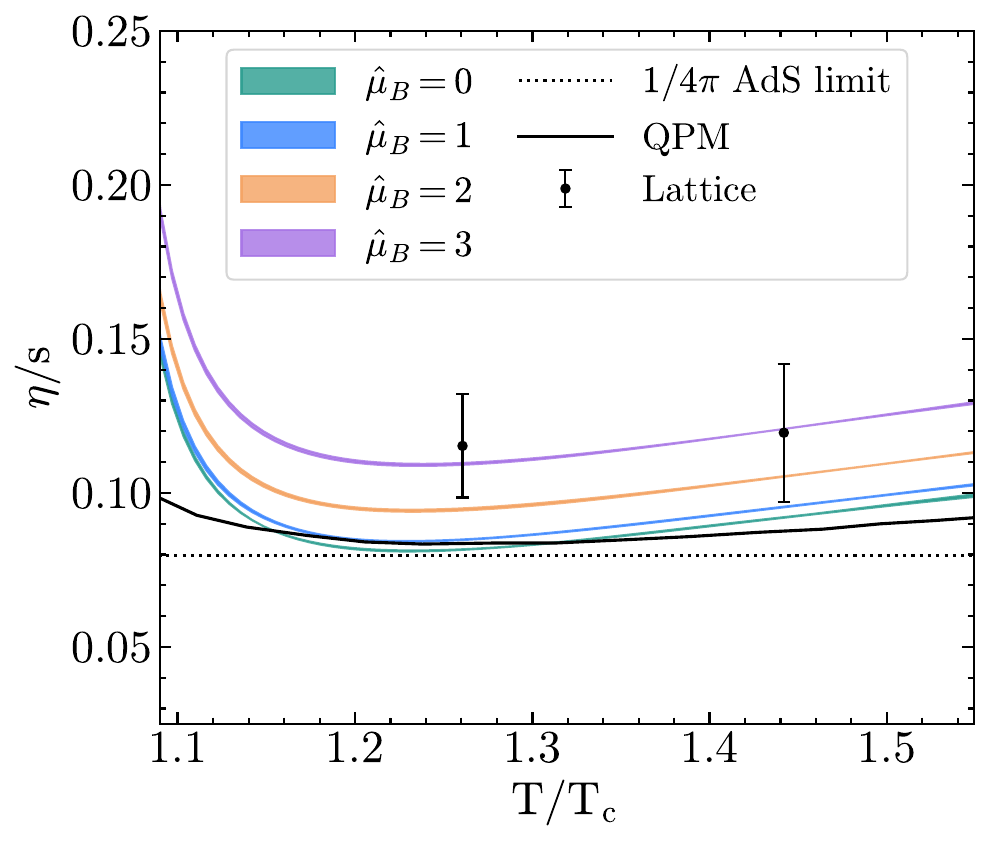}
    \includegraphics[width=0.45\textwidth]{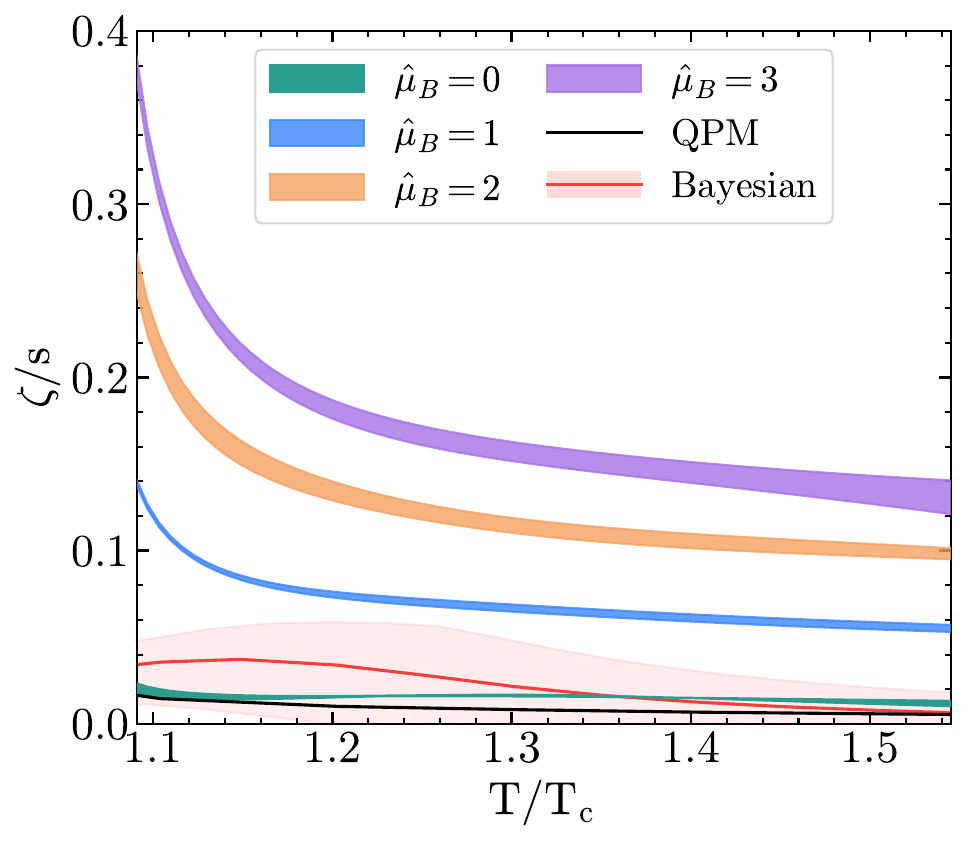}
    \caption{(Upper) Shear viscosity to entropy density ratio $\eta/s$ and (lower) bulk viscosity to entropy density ratio $\zeta/s$ as functions of $\TbyTc$ at different $\mucapB$ values. In the top panel, lQCD results~\cite{Haas:2013hpa} and the AdS/CFT bound $1/4\pi$~\cite{Kovtun:2004de} are shown for comparison. In the bottom panel, the red band represents the Bayesian estimate~\cite{Bernhard:2019bmu}, and the black line corresponds to the QPM result~\cite{Plumari:2011mk}.}
    \label{Fig:viscosity}
\end{figure}

Figure~\ref{Fig:cs2cvt3} shows the speed of sound squared $c_s^2$ and the scaled specific heat $C_V/T^3$ as functions of temperature. At $\hat{\mu}_{\mathrm{B}} = 0$, $c_s^2$ is in  agreement with lQCD estimates~\cite{Khvorostukhin:2010aj} and the experimental value reported by the CMS Collaboration~\cite{CMS:2024sgx}. With increasing temperature, $c_s^2$ rises monotonically and saturates toward the Stefan--Boltzmann limit of $1/3$~\cite{LandauLifshitz}, indicating the gradual restoration of conformal symmetry in the high-temperature perturbative regime. The enhancement of $c_s^2$  with increasing $\hat{\mu}_{\mathrm{B}}$ reflects the weakening of the effective interaction strength of the medium, consistent with the reduction of the quasi-particle thermal masses at finite baryon chemical potential. The scaled specific heat $C_V/T^3$ increases systematically with both temperature and $\hat{\mu}_{\mathrm{B}}$. The growth with $\hat{\mu}_{\mathrm{B}}$ is a direct consequence of the increase in the total energy density of the medium at finite baryon density, while the saturation of $C_V/T^3$ at high temperatures reflects the approach to the Stefan--Boltzmann regime, where the energy density scales as $\epsilon \sim T^4$, rendering $C_V/T^3$  approximately constant. The results are in quantitative agreement with available lQCD calculations~\cite{HotQCD:2014kol} across the full temperature range considered. \\

Further, Fig.~\ref{Fig:tau} exhibits the temperature dependence of the relaxation times for quarks and gluons, computed using Eq.~(\ref{Eq:tau}). We compare our results with  $\tau_q$ and $\tau_g$ relaxation times obtained within the QPM framework of Ref.~\cite{Plumari:2011mk} and with the quark relaxation time predicted by the NJL model in Ref.~\cite{Marty:2013ita}. We note that the quantitative difference between our results and those obtained from QPM in Ref.~\cite{Plumari:2011mk} arises from the choice of the parameter $k$ in Eq.~(\ref{Eq:tau}): Ref.~\cite{Plumari:2011mk} adopts $k=22$, whereas we use $k=18.5$, chosen specifically to enforce a minimum value of $\eta/s = 1/4\pi$ saturating the KSS bound.
The relaxation time $\tau$ decreases monotonically with increasing temperature, consistent with the growth of the number density of scattering centers, which shortens the mean free path and enhances the interaction rate of the quasi-particles. The quark relaxation time $\tau_q$ remains systematically larger than the gluon relaxation time $\tau_g$ across the full temperature range, reflecting the stronger color charge of gluons, characterized by the Casimir factor $C_A = 3$ compared to $C_F = 4/3$ for quarks, which leads to larger scattering cross-sections and consequently shorter relaxation times for gluons.\\

\begin{figure}
    \centering
    \includegraphics[width=0.45\textwidth]{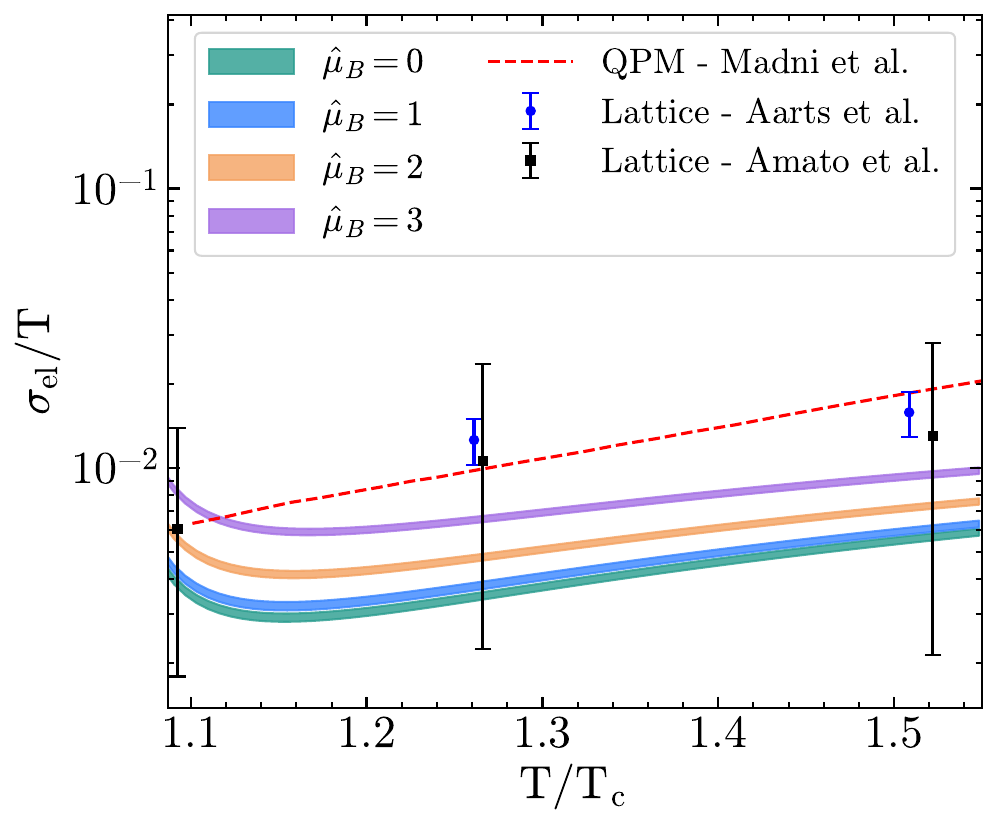}
    \includegraphics[width=0.45\textwidth]{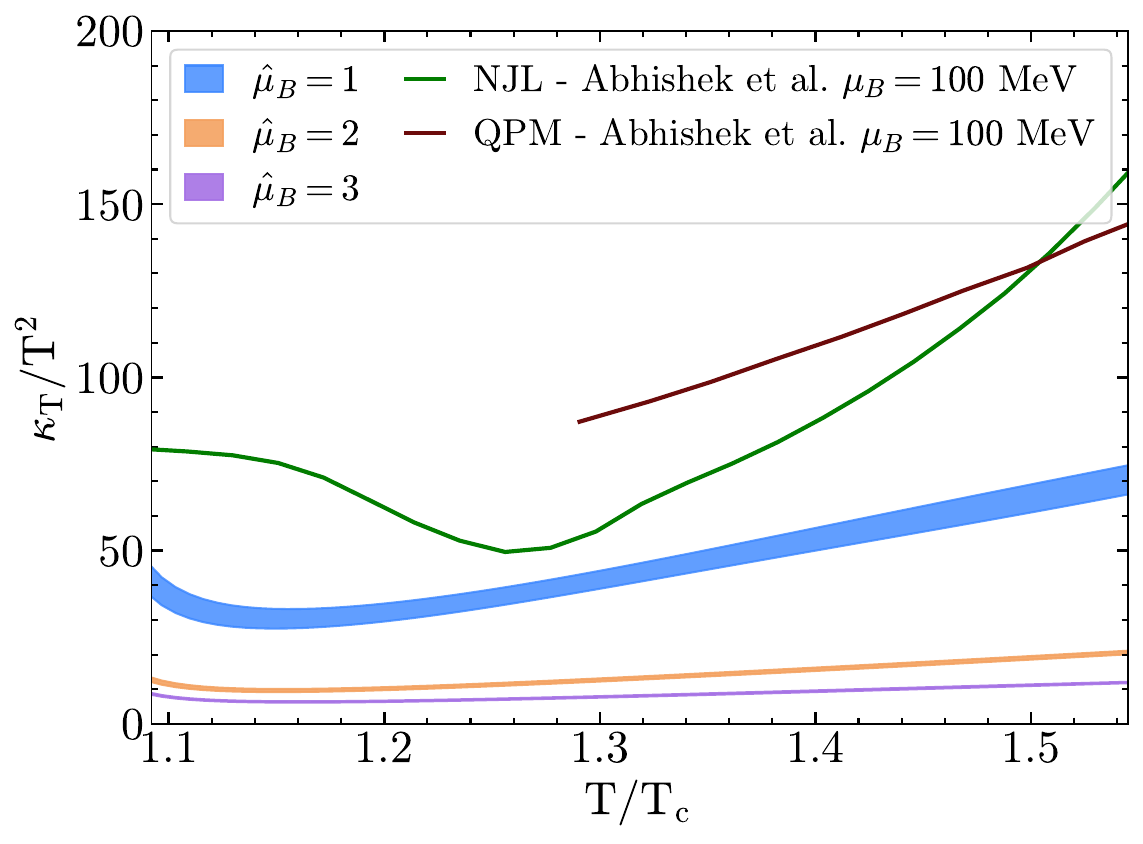}
    \caption{(Upper plot) The electrical conductivity as a function of scaled temperature $\mathrm{\sigma_{el}/T}$ at finite $\muB$ values compared to results from Madni et. al.~\cite{Madni:2024ubw} and lattice data~\cite{Aarts:2014nba, Amato:2013naa}. (Lower plot) The thermal conductivity as a function of temperature compared to results from Abhishek et. al. at finite $\mathrm{\muB~}$~\cite{Abhishek:2020wjm}.}
    \label{Fig:conductivity}
\end{figure}

The top panel of Fig.~\ref{Fig:viscosity} shows the shear viscosity to entropy density ratio $\eta/s$ as a function of $T/T_{\mathrm{c}}$. The ratio exhibits a minimum near $T \approx 1.2\,T_{\mathrm{c}}$, 
arising from the competing temperature dependences of the parton distribution functions and the relaxation time. Above this minimum, $\eta/s$ increases monotonically with temperature, reflecting the approach toward the weakly interacting perturbative regime. At $\hat{\mu}_{\mathrm{B}} = 0$, the results are in qualitative agreement with QPM estimates~\cite{Plumari:2011mk}. The results lie below the lQCD-based estimates of Ref.~\cite{Haas:2013hpa}, which carry relatively large systematic uncertainties, and remain above the 
KSS bound $\eta/s = 1/4\pi$~\cite{Kovtun:2004de}.\\

The bottom panel of Fig.~\ref{Fig:viscosity} shows the bulk viscosity to entropy density ratio $\zeta/s$ as a function of $T/T_{\mathrm{c}}$ for several values of $\hat{\mu}_{\mathrm{B}}$. At 
$\hat{\mu}_{\mathrm{B}} = 0$, $\zeta/s$ exhibits a weak temperature dependence and is consistent with QPM results~\cite{Plumari:2011mk}, as well as the $90\%$ credible region from a Bayesian 
analysis~\cite{Bernhard:2019bmu}. With increasing $\hat{\mu}_{\mathrm{B}}$, $\zeta/s$ is systematically enhanced, with a noticeably stronger temperature dependence emerging near $T_{\mathrm{c}}$. This behavior indicates that finite baryon chemical potential amplifies bulk viscous dissipation in the medium, particularly in the vicinity of the 
deconfinement transition, where non-perturbative effects are most pronounced.\\

The upper panel of Fig.~\ref{Fig:conductivity} illustrates the scaled electrical conductivity $\sigma_{\mathrm{el}}/T$ as a function of $T/T_{\mathrm{c}}$ for several values of $\hat{\mu}_{\mathrm{B}}$. At $\hat{\mu}_{\mathrm{B}} = 0$, $\sigma_{\mathrm{el}}/T$ exhibits a pronounced dip near $T_{\mathrm{c}}$, reflecting the strong non-perturbative interactions in the vicinity of the deconfinement transition, followed by a monotonic increase with temperature as the medium approaches the perturbative regime. The magnitude of $\sigma_{\mathrm{el}}/T$ increases systematically with $\hat{\mu}_{\mathrm{B}}$, indicating that finite baryon density enhances charge transport by increasing the number density of electrically charged carriers in the medium. At $\hat{\mu}_{\mathrm{B}} = 0$, the results are in qualitative agreement with kinetic theory estimates within the RTA using the Gribov--Zwanziger approach~\cite{Madni:2024ubw}, although quantitative differences remain. One important source of this difference is the relaxation time. In particular, the $\tau_{l/s}$ values reported in Ref.~\cite{Madni:2024ubw} are approximately $15$--$20$ fm/\textit{c}, whereas the quark relaxation times obtained in the present work are only about $0.5$--$1.5$ fm/\textit{c}$,$ comparable to the QPM results of Ref.~\cite{Plumari:2011mk}. Since the electrical conductivity scales approximately linearly with the relaxation time, a larger relaxation time naturally leads to larger conductivity values. The lQCD estimates~\cite{Aarts:2014nba, Amato:2013naa} remain somewhat higher than our results, which may reflect limitations of the relaxation-time approximation and the simplified relaxation dynamics adopted in the present framework.

The lower panel of Fig.~\ref{Fig:conductivity} shows the scaled thermal conductivity $\kappa_T/T^2$ as a function of $T/T_{\mathrm{c}}$  for several values of $\hat{\mu}_{\mathrm{B}}$. While $\kappa_T/T^2$  increases monotonically with temperature, it decreases systematically with increasing $\hat{\mu}_{\mathrm{B}}$. This behavior can be understood from Eq.~(\ref{eq:conductivity}), where the dominant contribution to $\kappa_T$ is governed by the factor $(E - b_i h)^2 \approx (b_i h)^2$, valid in the limit where the single-particle energy $E$ is small compared to the enthalpy per particle $h$. At finite $\hat{\mu}_{\mathrm{B}}$, the rapid growth of the net baryon density $n_{\mathrm{B}}$ reduces the enthalpy per baryon $h = (\epsilon + P)/n_{\mathrm{B}}$, which in turn suppresses $\kappa_T/T^2$. This indicates that a higher baryon chemical potential suppresses thermal transport — equivalently, heat diffusion — in the deconfined medium. The results are consistent with results from Abhishek et. al.~\cite{Abhishek:2020wjm} at finite $\hat{\mu}_{\mathrm{B}}$, with the overall temperature dependence in good agreement across all approaches.

\section{Summary}
\label{Summary}
In this work, we study the thermodynamic and transport properties of the QGP at finite baryon chemical potential within a deep-learning-assisted quasi-particle framework. We determine the effective temperature- and baryon chemical potential-dependent quasi-particle masses by training residual neural networks on lQCD results obtained via a Taylor-like expansion around vanishing chemical potential. The trained model acts as an effective emulator, enabling a consistent extension of the equation of state and transport coefficients into the finite $\muB$ regime, which remains challenging for first-principles calculations.
Using the extracted quasi-particle masses, we compute the thermodynamic observables through the grand-canonical partition function and observe a very good agreement with lQCD calculations across the explored $(T,\muB)$ range. The model captures the expected dependence of pressure, energy density, entropy density, and baryon number density on $\mucapB$, while maintaining a smooth and physically consistent behavior of the effective masses. We further estimate the transport properties of the medium within the relaxation-time approximation. We observe that the shear viscosity-to-entropy-density ratio exhibits a minimum near the transition region, followed by an increase at higher temperatures, reflecting the interplay between parton distributions and relaxation dynamics. The bulk viscosity shows an enhancement with increasing baryon chemical potential, particularly near $T_{\mathrm{c}}$, indicating stronger dissipative effects in dense QCD matter. Similarly, the electrical conductivity increases with $\muB$, while the thermal conductivity is suppressed due to the modification of the enthalpy per particle at finite baryon density. These trends highlight the sensitivity of transport coefficients to both temperature and chemical potential and are in qualitative agreement with existing phenomenological models and lattice-based estimates.\\

Overall, our results demonstrate that a deep-learning-assisted quasi-particle approach provides a flexible, data-driven framework for studying QCD matter at finite baryon density, while remaining consistent with lattice constraints at $\mu_{\mathrm{B}}=0$. However, while the thermodynamic quantities are well constrained through the learned quasi-particle masses, the determination of transport coefficients remains sensitive to the choice of relaxation time. A more fundamental, model-independent formulation of the relaxation dynamics is therefore essential for achieving a quantitatively reliable description of transport properties in the QGP.  In the future, with abundant lQCD data for the transport properties, one can train a similar ResNet model to extract the relaxation time from lattice data. Furthermore, extending this deep-learning approach to include heavy-flavor dynamics will be a crucial next step, as heavy quarks are known to provide significant theoretical contributions to the overall QGP viscosity \cite{Zaccone:2023ozw}.
These results provide a systematic framework for extending QCD thermodynamics and transport studies to finite baryon density and motivate further developments combining machine learning with first-principles constraints.

\section*{Acknowledgment}

K.G. acknowledges financial support from the Prime Minister's Research Fellowship (PMRF), Government of India. S.P. acknowledges the support from the Hungarian National Research, Development and Innovation Office (NKFIH) under the contract numbers NKFIH NKKP ADVANCED\_25-153456, 2025-1.1.5-NEMZ\_KI-2025-00005, 2024-1.2.5-TET-2024-00022, and the usage of Wigner Scientific Computing Laboratory (WSCLAB). M.Y.J. acknowledges Central China Normal University (CCNU), Wuhan, China, for its hospitality and financial support through a postdoctoral fellowship. R.S. acknowledges the DAE-DST, Government of India, funding under the mega-science project “Indian participation in the ALICE experiment at CERN” bearing Project No. SR/MF/PS-02/2021-IITI(E-37123). 

\appendix
\section{Model Training}
\label{appendix}

The data used in this work are sourced from WB lattice calculations at finite chemical potential for the QCD EoS~\cite{Borsanyi:2021sxv}.
The temperature- and chemical potential-dependent quasi-particle masses are represented by three distinct Residual Neural Networks (ResNets), Table~\ref{tab:nn_architecture} details the model architecture used in this work. This is structured similarly to the architecture used by Li \textit{et al.} in Ref.~\cite{Li:2025csc}. \par
\begin{table}[h]
\centering
\caption{Neural network architecture used to parametrize the temperature- and chemical potential-dependent quasi-particle masses. Three identical networks are employed to model the $u/d$ quark, the strange quark, and the gluon masses.}
\label{tab:nn_architecture}
\begin{ruledtabular}
\begin{tabular}{l c c}
\textbf{Component} & \textbf{Layer type} & \textbf{Dimensions} \\
\hline
Input & Fully connected & $2 \rightarrow 16$ \\
Expansion & Fully connected & $16 \rightarrow 32$ \\
Residual Block $\times 7$ & Fully connected & $32 \rightarrow 32$ \\
Compression & Fully connected & $32 \rightarrow 16$ \\
Output & Fully connected & $16 \rightarrow 1$ \\
\hline
Activation (hidden) & Swish & $x\,\sigma(x)$ \\
Activation (output) & Sigmoid & $2 \times \sigma(x)$ \\
\end{tabular}
\end{ruledtabular}
\end{table}
The model hyperparameters are summarized in Table~\ref{tab:training_hyperparameters}. To reduce the overall training time compared to previous studies, the total number of training epochs is reduced. An initial learning rate of $10^{-3}$ is adopted with a decay factor of $0.92$. The mean absolute error (MAE) is used as the loss function, and the model is optimized using the AdamW algorithm. The dataset is randomly split into $80\%$ for training and $20\%$ for validation in order to monitor overfitting and evaluate the generalization performance of the model.
\begin{figure}[h]
    \centering
    \includegraphics[width=.9\linewidth]{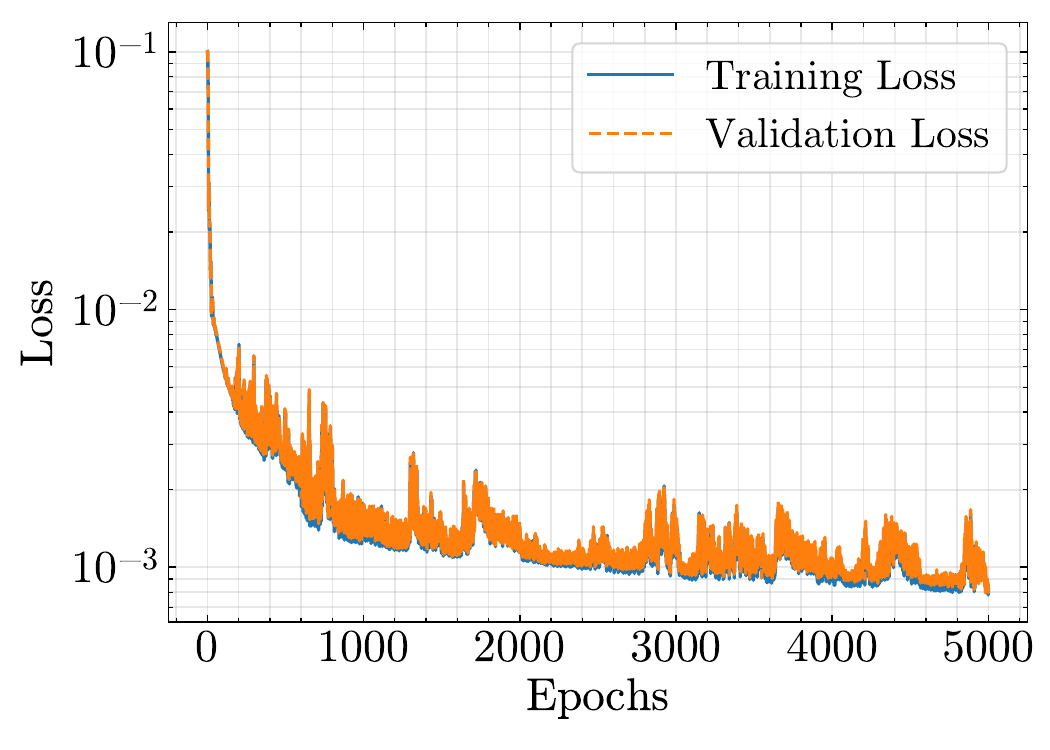}
    \caption{The training-validation loss curve represents the loss throughout the training of the model. The solid blue line indicate the training loss, and the dashed orange line indicate the validation loss.}
    \label{fig:lossvalid}
\end{figure}
Figure~\ref{fig:lossvalid} shows the training and validation loss curves. After approximately $3000$ epochs, the loss saturates, indicating convergence toward a local minimum. During training, the model parameters are updated using the AdamW optimizer, and the parameter set corresponding to the lowest validation loss is retained. The close agreement between the training and validation curves indicates that the model provides a good fit to the data without significant overfitting.

\begin{figure}[h]
    \centering
    \includegraphics[width=.9\linewidth]{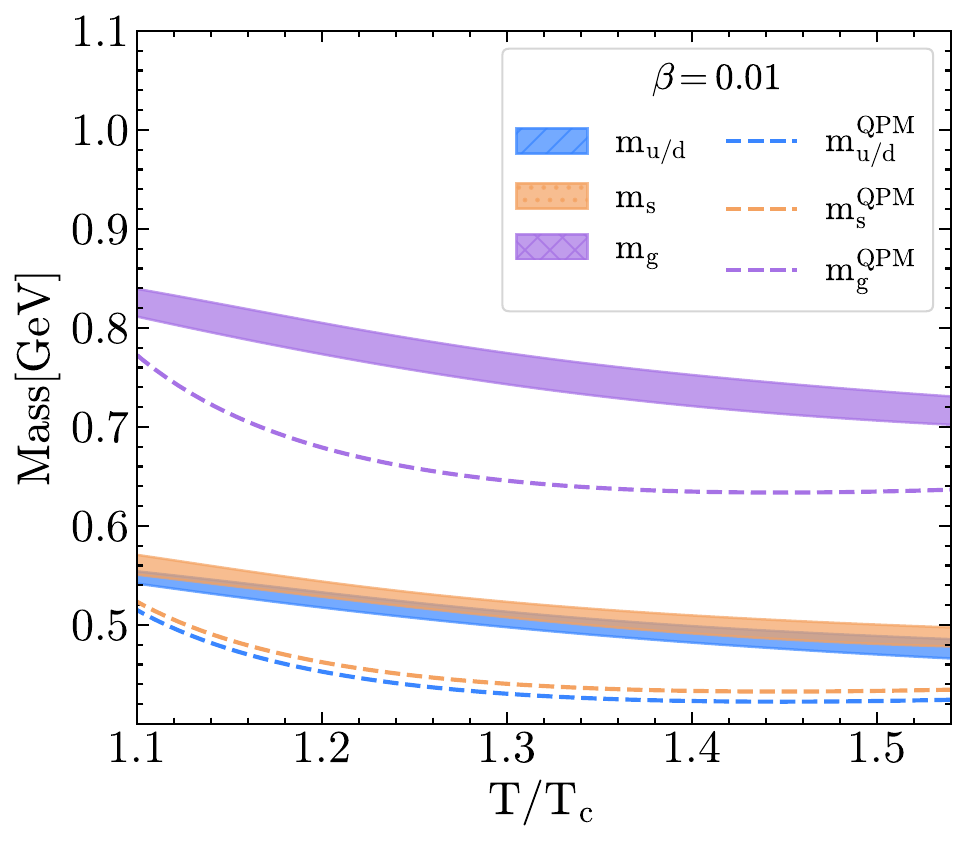}  
    \caption{The plot represents the quasiparticle masses predicted by the model as a function of $\mathrm{T}$ at $\mathrm{\muB~ = 0}$. The masses $\mathrm{m_{u/d}}$, $\mathrm{m_s}$, and $\mathrm{m_g}$ are given by the blue, orange, and purple bands, respectively. The dashed lines represent the masses obtained using relation (\ref{eq:mass_formulae}).}
    \label{fig:mass0mub}
\end{figure}

\begin{figure}[h]
    \centering
    \includegraphics[width=.8\linewidth]{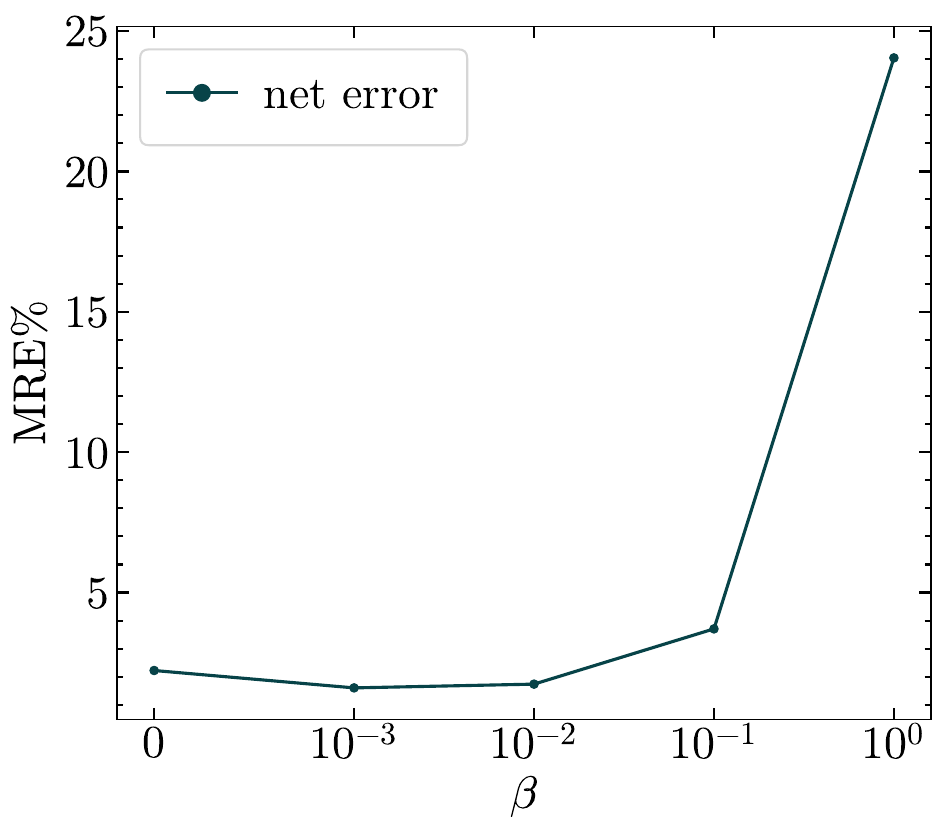}
    \caption{Mean percentage error as a function of the mass-regularisation scaling term $\beta$.}
    \label{fig:errorbeta}
\end{figure}

Figure~\ref{fig:mass0mub} illustrates the effective masses of the quasi-particles as functions of temperature computed at $\muB~=0$. We observe that the effective masses decrease monotonically as temperature increases. We compare our results with QPM masses calculated from Ref.~\cite{Plumari:2011mk}, and observe that the quasi-particle masses predicted by DLQPM are slightly higher; however, it follows a similar monotonically decreasing trend with temperature.
Moreover, in Fig.~\ref{fig:errorbeta}, we plot the mean relative error (MRE) percentage as a function of $\beta$ to validate the selection of $\beta$ values in Eq.~\ref{Eq:mass_reg}. We estimate the MRE as,
{\footnotesize
\begin{equation}
\begin{aligned}
\mathrm{MRE} = \frac{1}{5} \Bigg( 
&\left|\frac{(P)_{\mathrm{pred}} - (P)_{\mathrm{WB}}}{(P)_{\mathrm{WB}}}\right|
+ \left|\frac{(s)_{\mathrm{pred}} - (s)_{\mathrm{WB}}}{(s)_{\mathrm{WB}}}\right| \\
+ &\left|\frac{(\Delta)_{\mathrm{pred}} - (\Delta)_{\mathrm{WB}}}{(\Delta)_{\mathrm{WB}}}\right|
+ \left|\frac{(\varepsilon)_{\mathrm{pred}} - (\varepsilon)_{\mathrm{WB}}}{(\varepsilon)_{\mathrm{WB}}}\right| \\
+ &\left|\frac{(n_B)_{\mathrm{pred}} - (n_B)_{\mathrm{WB}}}{(n_B)_{\mathrm{WB}}}\right|
\Bigg)
\end{aligned}
\label{eq:mape}
\end{equation}
}
As shown in Fig.~\ref{fig:errorbeta}, the prediction error is minimized for $\beta \approx 0.01$ and $0.001$, indicating an optimal balance between reproducing the lattice thermodynamics and constraining the quasi-particle masses. For $\beta \geq 0.1$, the mass constraints are too dominant, which leads to a noticeable disagreement between DLQPM-predicted thermodynamic observables and WB results.

Once the training is completed, the model can reliably predict the effective masses of the quasi-particle, reproducing its training data and predicting the transport properties of hot QCD matter at finite baryon chemical potential. 

\newpage

 
\end{document}